\documentstyle[11pt,ska,twoside,psfig]{article}
\begin{document}
\title{Extra-solar planets with SKA}
\author{Alan J. Penny}
\affil{ Rutherford Appleton Laboratory, Chilton, Didcot OX11 0QX, 
\newline United Kingdom}

 
\begin{abstract} 

SKA would have distinct capabilities for studies associated with
extra-solar planets, their formation and properties.

\end{abstract}

\section{Introduction}

This paper discusses the use of SKA in the study of extra-solar
planets. It assumes a SKA with a collecting area of a square
kilometre, but with many of its antennae spread over some hundreds of
kilometres, giving an angular resolution of 0.1 arcsec at 20 cm. The
spectral range assumed is from 2 cm to 1 metre, with the short
wavelength limit determined in part by the ALMA performance.

\section{Present extra-solar planet knowledge}

At present some hundred Jupiter-mass planets have been detected
orbiting nearby stars by the radial velocity method. This method
gives the orbital radius, the orbit eccentricity and a lower mass
limit. Statistically, the true planet masses should average some 30\%
greater than these lower limits. The mass lower limits are in the
range of 0.1 to 15 Mjupiter and the orbital radii are the range 0.04
AU to 6 AU. Orbital eccentricities are in the range 0.0 to 0.7. The
limits of and distributions within these ranges are affected by
observational selection effects. One planet has been detected to
transit its parent star, so a precise mass and radius can be derived.
For this planet, the presence of sodium in the atmosphere has been
detected through transit spectroscopy. Another planet has had its
astrometric motion measured by the HST Fine Guidance Sensor, giving
the mass. A few stars have two or three planets. The parent stars
tend to be metal-rich. Two other transiting planets have been claimed
but with limited data. Microlensing searches have not yet found any
planets. Theories of planet formation are still in dispute both for
giant and terrestrial planets, as are the details of the migration
processes that produce the close-in Jupiters. However theoretical
calculations of the masses, sizes and compositions of the close-in
Jupiters are giving reasonable good agreement with observations.

.
\section{Extra-solar planet knowledge when SKA begins}

Over the next decade or so ground work will become much more
powerful. The radial velocity searches should find many more planets,
and be able to detect Neptune-mass planets, and Jupiters in 12 year
orbits. Ground-transit searches should also detect many planets, a
few bright enough to permit more detailed studies. Micro-lensing
searches will probably have detected planets, and if some plans come
to fruition, be able to detect Earth-mass planets. Direct imaging
detection through AO should detect some bright nearby Jupiters. If
the ELT 30-metre to 100-metre telescope come about, there are claims
that they will be able detect Earths.

Over the same timescale there will be a revolution in space work.
Table 1 shows the presently funded (apart from Darwin/TPF) missions
which are either directly for extrasolar planet work or will be able
to study them.

\vspace*{1em}
\begin{table}
\begin{tabular}{llll}
Planet Type    &   Mission           &    Method       &  Date  \\
               &                          &            &            \\
Jupiter            &   MOST            &    reflection    &  2003  \\
Large Earth    &   COROT         &    transit         &  2005  \\
Earth              &   Kepler            &    transit         &  2007  \\
Earth              &   Eddington      &     transit        &  2007  \\
Large Earth    &   SIM               &    astrometry  &  2010  \\
Jupiter            &   JWST             &   imaging      &  2011  \\
Jupiter            &   GAIA             &    astrometry  &  2012  \\ 
Earth              &    Darwin/TPF  &    imaging/spectra & 2015 \\ 
\end{tabular}
\caption{Future planet-related space missions}
\end{table}
\vspace*{1em}

These missions will detect many thousands of planets and determine
the frequency of occurrence and properties of many types of planets
from Earth size to Jupiter size.

\section{Other planet-related work}

\subsection{Protoplanetary disks}

Much work is being done on the early stages of formation when the
planets form out the protoplanetary disk, both observationally 
and theoretically (see e.g. Boss, 2001 and Mayer et al, 2002),
but many uncertainties remain. For example, it is not understood how
planetesimals coalesce to form the terrestrial planets. Optical,
millimetre and sub-millimetre imaging and spectroscopy have given
information on some protoplanetary disks on the 10s of AU scales.

\subsection{Circumstellar disks}

Around certain old stars, there exist circumstellar disks. These are
of interest for extrasolar planets because they provide information
on the outer regions of planetary systems for objects similar to
Kuiper Belt objects, and structures in these disks can be interpreted
as due to the influence of interior planets. For a review see
Zuckerman (2001).

\subsection{Planetary non-thermal emission}

Jupiter emits coherent cyclotron radiation from electrons travelling
along field lines into the auroral regions and there is decametric
radiation from the Io-Jupiter interaction. There was water maser
emission from Jupiter when the Shoemaker-Levy 9 comet hit Jupiter.

\subsection{Prebiotic molecules in the ISM}

Prebiotic molecules in the interstellar medium (see e.g. Hjalmarson
et al, 2001) are of interest for the composition of extrasolar
planets, as these molecules will have been included in the chemical
processes of their formation and in subsequent delivery by comets.
They may also be relevant to the origins of life on terrestrial
planets. Many long-chain molecules have been discovered so far, with
H2C6 the longest cumulene currently known.

\subsection{Pulsar planets}

Four planet-mass objects have been detected orbiting two pulsars (see
e.g. Konacki et al , 2000). Two of these are Earth mass, and one is
Moon mass. The detection method is to look for variations in the
pulse epochs due to the Keplerian orbit of the pulsar. No widely accepted
theory of the formation or nature of these objects has been made.

\section{SKA work on extrasolar planets}

The prospects for SKA have to be seen in the light of the work that
will be done with ground and space investigations over the next
decade or so,

\subsection{Protoplanetary disks}

These disks become optically thick due to dust obscuration in the
central regions. SKA continuum and line (such as NH3) observations
will give structure and dynamic imaging of the dense molecular gas
associated with star formation cores on sub-AU scales.

\subsection{Circumstellar disks}

The sensitivity of SKA will permit the radio-wavelength thermal
emission from circumstellar disks to be detected. This will
complement the millimetre and sub- millimetre work by ALMA.

\subsection{Astrometric planet detection}

The milliarcsec capability in astrometry of SKA, when in the dilute
array configuration, for nearby stars will be inferior to the
performance of the SIM and GAIA space missions, both in accuracy and
number of targets, and is thus unlikely to be of significant value.

\subsection{Direct detection}

\subsubsection{Thermal emission}

The thermal emission form a close-in Jupiter at 1000K at a distance
of 1 pc at 1cm wavelength would be 500 nJy, compared to the 300
{$\mu$}Jy from the star (de Pater, 2002). The planet spectral shape
thus might just be distinguishable from the star spectral shape. The
angular separation would be 0.05 arcsec, compared with SKA resolution
of 2 arcsec if configured as a large array, so they could not be
resolved.

\subsubsection{Non-thermal emission}

de Pater (2002) has calculated the possible emission from some known
extrasolar planets from coherent cyclotron radiation from electrons
travelling along field lines into the auroral regions., and finds
fluxes of the order of mJy at 40-300 MHz, easily detectable by SKA.
The fluxes may be much higher during active periods in the stars. de
Pater also points out the possibility of detecting decametric
radiation, similar to the Io-Jupiter interaction, from a
planet-satellite or a star-planet interaction.

Recently, there has been a claimed detection (Cosmovici, 2002), but
not confirmed (Butler et al, 2002), of emission from three extrasolar
planet systems of water maser emission. This has been postulated as
coming from the same pumping process that was observed when the
Shoemaker-Levey 9 comet hit Jupiter. SKA would be able to search for
such radiation with great sensitivity.

\subsection{Prebiotic molecules in the ISM}

To reach longer chain molecules than presently detectable, increased
sensitivity will be essential. As such molecules will have larger
moments of inertia, their emission will be in the centimetre range,
and as there are many transition modes, the energy for each
transition will be less. Calculations predict a lower abundance for
these larger molecules. Further, it will be essential to look for
these molecules in dense regions and the better sensitivity and
angular resolution of SKA will permit more powerful searches for
small sources, and also by mapping their surroundings understand
their evolutionary phase.

\subsection{Pulsar planets}

Resolving the puzzle of the origin and nature of the planet-mass
objects orbiting two pulsars will depend on detecting and studying
more of them. The sensitivity of SKA will permit greatly enhanced
pulsar searches to be made.

\acknowledgements

Much of the discussion in this paper was informed by the SKA Science
Case available at the SKA telescope web site
\footnote{http://www.skatelescope.org/ska\_science.shtml} and by other SKA
project documents.

\clearpage 
.
\end{document}